\documentclass[useAMS,usenatbib]{mn2e}

\usepackage{times}
\usepackage{graphics,epsfig}
\usepackage{graphicx}
\usepackage{amsmath}
\usepackage{amssymb}
\usepackage{rotating}
\usepackage{ulem}
\usepackage{float}
\usepackage{booktabs}
\usepackage{color}

\newcommand{\kms}{km~s$^{-1}$}

\newcommand{\ha}{\ensuremath{{\rm H}\alpha}}
\newcommand{\ms}{\ensuremath{\rm M_{\odot}}}

\newcommand{\um}{\ensuremath{\rm \mu m}}
\newcommand{\sgas}{$\Sigma_{\rm gas}$}
\newcommand{\shi}{$\Sigma_{\rm gas, atomic}$}
\newcommand{\shm}{$\Sigma_{\rm H_2}$}
\newcommand{\ssfr}{$\Sigma_{\rm SFR}$}

\title[K--S relation in HI dominated regime]{The spatially resolved Kennicutt-Schmidt relation in the HI dominated regions of spiral and dwarf irregular galaxies}

\author[Roychowdhury et al.]{Sambit Roychowdhury,$^{1}$\thanks{E-mail: sambit@mpa-garching.mpg.de} Mei-Ling Huang,$^{1}$ Guinevere Kauffmann,$^{1}$ Jing Wang$^{2}$ 
\newauthor and Jayaram N. Chengalur$^{3}$ \\
       \\ 
       $^{1}$Max-Planck-Institut f\"{u}r Astrophysik, Karl-Schwarzschild-Str. 1, 85748 Garching, Germany\\
       $^{2}$Australia Telescope National Facility, CSIRO, P.O. Box 76, Epping, NSW 1710, Australia\\
       $^{3}$NCRA-TIFR, Post Bag 3, Ganeshkhind, Pune 411 007, India}

\begin{document}
\date{}

\pagerange{\pageref{firstpage}--\pageref{lastpage}} \pubyear{}

\maketitle

\label{firstpage}

\begin{abstract}
We study the Kennicutt–-Schmidt relation between average star formation rate and average cold gas surface density
in the HI dominated ISM of nearby spiral and dwarf irregular galaxies. We divide galaxies
into grid cells varying from sub-kpc to tens of kpc in
size. Grid-cell measurements of low SFRs using H$\alpha$~emission can be biased and
scatter may be introduced because of non-uniform sampling of the IMF
or because of stochastically varying star formation. In order to alleviate these issues, 
we use far-ultraviolet emission to trace SFR, and we sum up the fluxes 
from different bins with the same gas surface density to calculate the {\it average} \ssfr\
at a given value of $\Sigma_{gas}$. We study the resulting Kennicutt–-Schmidt
relation in 400 pc, 1 kpc and 10 kpc scale grids in nearby massive spirals and in 400 pc scale grids 
in nearby faint dwarf irregulars. We find a relation with a power law slope of 
1.5 in the HI-dominated regions for both kinds of galaxies. 
The relation is offset towards longer gas consumption timescales compared
to the molecular hydrogen dominated centres of spirals, but the offset is
an order-of-magnitude less than that quoted by earlier studies.  
Our results lead to the surprising conclusion that conversion of gas to stars is independent of metallicity in the HI dominated regions of star-forming galaxies.
Our observed relations are better fit by a model of star formation 
based on thermal and hydrostatic equilibrium in the ISM, in which feedback driven turbulence sets the thermal pressure. 
\end{abstract}


\section{Introduction}
\label{sec:int}

How gas is converted into stars is an important question crucial to our understanding of galaxy formation and evolution.
An empirical way of quantifying this  was given by \citet{s59} and extended by \citet{ken98}.
The Kennicutt--Schmidt (K--S) relation is a power law relation between \ssfr\ and \sgas\ 
averaged over the star-forming disks of nearby galaxies.
Insight into the physical processes regulating star formation has 
been sought by trying to model the K--S relation \citep{kru09,ost10} using physical principles as well as empirical clues.

Only recently have surveys of nearby galaxies reached sufficient sensitivities at all 
requisite wavelengths to study the K--S relation on sub-kpc scales.
Examples include the study of nearby spiral galaxies by \citet{big08}.
In a similar study, \citet{ler13}  showed that a linear K--S relation exists 
between the surface density of molecular gas and \ssfr.
This is not unexpected as the molecular phase of hydrogen traces the molecular clouds where star formation
actually occurs. 
A linear K--S relation between \shm\ and \ssfr\ has also been shown to hold 
in the atomic hydrogen dominated outer discs of nearby spirals \citep{sch11}.
\citet{big10} showed that in the very outskirts of spiral discs (outside R$_{\rm 25}$), where molecular hydrogen cannot be detected  and atomic hydrogen (HI) dominates, 
there is a clear relationship between FUV and HI column density, with the
star formation efficiency  increasing with increasing HI column. The inferred
gas consumption timescales in these regions were well above a Hubble time ($\sim 10^{11}$ years).
Similar results  were found in recent resolved sub-kpc studies of HI
and star formation in dwarf galaxies by \citet{roy09,roy11}.

One crucial factor to take into account when  determining the relation between gas and star formation 
in the HI dominated regime is the low star formation rate surface densities in this regime.
Very deep imaging data is required to measure  fluxes accurately. In addition,
since direct tracers of star formation primarily trace young high mass stars 
($\gtrsim$ 16 \ms\ for \ha, $\gtrsim$ 6 \ms\ for FUV), 
the high mass end of the initial mass function (IMF) will be sampled stochastically 
for low SFRs resulting in considerable fluctuation in the luminosity of the tracer 
across an ensemble of systems with the same SFR.
Bursty SF histories will also result in errors when using calibrations which 
assume continuous star formation during the previous Myr or so.
\citet{daS14} have simulated the effects of stochasticity and bursty SF history on the measurement of SFRs using different tracers. They show that that a bursty SF history leads to the peak of the distribution of SFRs measured for an ensemble of simulated systems to be offset on the lower side compared to the true SFR put in as input, and this offset increases with decreasing SFR. Stochasticity in the sampling of the IMF leads to the scatter in measured SFRs increasing with decreasing SFR.
In a recent study, \citet{roy14} computed a relation between
SFR and atomic gas surface density in nearby dwarf irregulars {\it averaged} over their star-forming discs, and
found that the \shi\ based K--S relation is not steep, but almost linear.
The implied efficiency of star formation was low in HI dominated dwarf galaxies, 
but it did not decrease as a function of HI surface density as found by  \citet{big10}. 
A more systematic study of the relation between \ssfr\ and \sgas\ in the HI dominated 
regime is therefore now in order.
Here we study the relation between \ssfr\ and \sgas\ in nearby $\sim L_*$ spirals and 
dwarf galaxies over different scales, carefully accounting for how we measure low SFRs.
We start with kpc sized regions and zoom in further to study 
trends such as changes in scatter in the \ssfr\ -- \sgas\ plane.
We also study how the K--S relation changes for few tens of kpc sized
regions in a sample of $\sim L_*$ spirals with very large HI-to-optical size ratios.

\section {The Samples}
\label{sec:samp}

\subsection {THINGS spirals}
\label{ssec:sampT}

The first sample is of spiral galaxies from The HI Nearby Galaxy Survey \citep[THINGS,][]{wal08}.
THINGS is a high spatial and velocity resolution interferometric survey of HI in 34 
nearby galaxies using the NRAO Very Large Array (VLA).
The sample spans a wide range in luminosity and mass, and included a handful of dwarf galaxies.
The THINGS galaxies used in the present study have 
CO detections from the HERA CO Line Extragalactic Survey \citep[HERACLES][]{ler09},
{\it Spitzer} 24 \um\ data, and low inclination angles. This is the same
sample used by \citet{ler13} to derive the linear K--S relation between \ssfr~and \shm.
Our sample is also part of the study of the very outskirts of spirals by \citep{big10}.
All galaxies have {\it GALEX} FUV observations which in 
combination with {\it Spitzer} 24 \um\ data is used to estimate their SFRs. 
We exclude one galaxy (NGC 3184) from the \citet{ler13} sample whose 
{\it GALEX} observations are from the low exposure time All-sky Imaging Survey (AIS) 
rather than being from the Medium Imaging Survey (MIS) or Guest Investigator data.
Another galaxy (NGC 2841) was excluded for not having CO data sensitive enough to trace the full disk, especially the central regions where there is high star formation but a prominent hole in the HI distribution. Tracing the CO in the central regions of our sample galaxies is important because as mentioned in Section~\ref{ssec:gas}, we use CO emission to exclude regions where HI does not dominate the gas in the ISM.

The final sample of THINGS spiral galaxies are listed in  Table~\ref{tab:sampT}.
Columns (1) gives the galaxy name, (2) the mass in stars \citep[from][]{ler13}, (3) the mass in HI \citep[from][]{wal08}, (4) the distance to the galaxy taken from \citet{ken11} for all galaxies except NGC 5194 \citep{wal08}.
Column (5) gives the resolution in kpc corresponding to the CO resolution of $\sim$13 arcsec.
We require HI maps at linear resolutions of $\sim$400 pc (see Section~\ref{sec:ana}), 
and seven of the galaxies had maps with average beam sizes comparable to 400 pc depending on the weighting scheme chosen: natural (NA) or robust (RO), and these are listed in column (6).

\begin{table}
\caption{THINGS spiral galaxy sample}
\label{tab:sampT}
\begin{tabular}{|lccccc|}
\hline
Galaxy&M$_{*}$&M$_{\rm HI}$&D&CO res.&HI cube\\
      &(10$^{\rm 10}$~M$_{\odot}$)&(10$^{\rm 8}$~M$_{\odot}$)&(Mpc)&(kpc)&400 pc\\
\hline
NGC 0628&1.0& 38.0& 7.2&0.46&NA\\
NGC 2903&2.5& 43.5& 8.9&0.57&  \\
NGC 3198&1.0&101.7&14.1&0.91&  \\
NGC 3351&1.3& 11.9& 9.3&0.60&NA\\
NGC 3521&5.0& 80.2&11.2&0.72&RO\\
NGC 3627&3.2& 8.18& 9.4&0.61&NA\\
NGC 4736&1.6& 4.00& 4.7&0.30&  \\
NGC 5055&3.2& 91.0& 7.9&0.51&  \\
NGC 5194&3.2& 25.4& 7.9&0.52&NA\\
NGC 5457&2.5&141.7& 6.7&0.43&  \\
NGC 7331&6.3& 91.3&14.5&0.94&NA\\
\hline
\end{tabular}
\end{table}

\subsection {FIGGS dwarf irregulars}
\label{ssec:sampF}

Our sample of dwarf irregular galaxies is derived from the Faint Irregular galaxy GMRT Survey \citep[FIGGS,][]{beg08}.
FIGGS is a high spatial and velocity resolution interferometric survey of HI in extremely faint (M$_{\rm B}~>~-$14.5) gas rich (single dish HI flux $>$ 1 Jy \kms) dwarf galaxies within a volume of $\sim$11 Mpc.
From the FIGGS sample we choose the galaxies which have archival {\it GALEX} far-ultraviolet (FUV) data,
which can be used for estimating their SFRs.
Once again  galaxies observed as part of the low sensitivity AIS are excluded. 
We also exclude galaxies with HI distributions that are morphologically offset from their stellar disc as found by \citet{roy14}.
The FIGGS galaxies are highly neutral hydrogen dominated and have a typical gas--to--total baryon ratio of $\sim$ 0.7, and have extended HI discs with the ratio of HI-to-optical diameter $\sim$2 \citep{beg08}.
The FIGGS galaxies have no detectable CO emission which can be used to trace molecular hydrogen. 

The properties of the FIGGS galaxies used in this study are given in Table~\ref{tab:sampF}.
Column (1) gives the galaxy name.
Some of the sample galaxies have available archival data from {\it Spitzer} based surveys of local galaxies.
These galaxies are indicated with a `S' in superscript after their names, followed by the reference to the original survey papers.
Column (2) gives the mass in stars \citep[see][for how the values were calculated]{beg08b}, (3) the mass in HI \citep[from][]{beg08}, (4) the distances to the galaxies \citep[from][]{kar13}, (5) the metallicity of the galaxy.
For only a handful of the sample galaxies (indicated with the reference number as superscript) abundance measurements are available, from which the metallicities given in column (5) are calculated \citep[see][for details]{roy14}.
For the rest of the galaxies the metallicity tabulated is an estimate using the luminosity (M$_{\rm B}$) -- metallicity relation for dIs from \citet{ekt10}.
These metallicity values are used to provide a correction factor to the measured SFR with the method described in Section~\ref{ssec:est}.

\begin{table}
\caption{FIGGS dwarf galaxy sample}
\label{tab:sampF}
\begin{tabular}{|lcccc|}
\hline
Galaxy&M$_{*}$&M$_{\rm HI}$&D&Z\\
      &(10$^{\rm 6}$~M$_{\odot}$)&(10$^{\rm 6}$~M$_{\odot}$)&(Mpc)&(Z$_{\odot}$)\\
\hline
And IV                   &    &205.19&6.3~&0.06~~~~\\
DDO 226$^{S,1}$          &35.5& 25.95&4.9~&0.12~~~~\\
UGC 685$^{S,1}$          &61.3& 56.15&4.5~&0.20$^3$\\
KK 65                    &74.3& 34.38&7.62&0.12~~~~\\
UGC 4115                 &42.3&285.53&7.5~&0.12~~~~\\
UGC 4459$^{S,1}$         &20.1& 64.2 &3.56&0.13$^3$\\
UGC 5456$^{S,1}$         &64.6& 58.96&5.6~&0.16~~~~\\
UGC 6456$^{S,2}$         &43.0& 43.89&4.3~&0.10$^4$\\
NGC 3741$^{S,1}$         &14.0&130.0 &3.0~&0.09$^3$\\
UGC 7242$^{S,1}$         &31.2& 45.75&5.4~&0.11~~~~\\
CGCG 269$-$049$^{S,1}$   &10.9& 26.4 &4.9~&0.05$^3$\\
UGC 7298                 &10.8& 21.6 &4.21&0.06~~~~\\
DDO 125$^{S,1}$          &69.6& 31.87&2.5~&0.12~~~~\\
UGC 7605$^{S,1}$         &27.7& 22.29&4.43&0.10~~~~\\
UGC 8215                 & 5.5& 21.41&4.5~&0.06~~~~\\
DDO 167                  &11.5& 14.51&4.2~&0.07~~~~\\
UGC 8638$^{S,1}$         &38.4& 13.76&4.27&0.10~~~~\\
DDO 181$^{S,1}$          &15.3& 27.55&3.1~&0.14$^3$\\
DDO 183$^{S,1}$          & 9.6& 25.90&3.24&0.09~~~~\\
UGC 8833                 & 7.8& 15.16&3.2~&0.07~~~~\\
DDO 187$^{S,1}$          &10.1& 16.30&2.5~&0.11$^3$\\
KKH 98                   & 2.8&  6.46&2.5~&0.04~~~~\\
\hline
\end{tabular}
\begin{flushleft}
References-- 1: \citet{dal09}; 2: \citet{eng08}; 3: \citet{mar10}; 4: \citet{mou06}.\\
\end{flushleft}
\end{table}

\subsection {Bluedisk spirals}
\label{ssec:sampB}

The final sample used in this study comes from a survey of 
11~$>~{\rm log~(M}_*/{\rm M}_{\odot})~>$~10, 0.01~$<$~z~$<$~0.03 galaxies with unusually 
high HI mass fractions -- 0.6 dex higher than the median relation between M(HI)/M$_*$ and M$_*$ found by \citet{cat10}.
Since these galaxies were chosen based on the bluer $g~-~i$ colour in the outer 
optical disc compared to the inner optical disc \citep[see][for the details of the survey]{wan13}, 
they are termed `Bluedisks'.
These galaxies have HI discs extending much beyond their optical discs.
An equivalently large sample of spirals with stellar masses and mass surface densities comparable to the primary `Bluedisk' sample were also observed in HI, and these form the `control' galaxy sample.
We exclude the galaxies identified as multi-source systems from our sample.

The `Bluedisk' and 'control' galaxies do not yet have CO measurements, but here we only focus on the 
HI-dominated regime by excluding the central regions (within R25) of the galaxies
where \sgas\ is likely to be dominated by molecular hydrogen.
{\it GALEX} FUV data combined with {\it WISE} 22 \um\ data is used as the main tracer of SFR. 
For 27 of 42 sample galaxies, the source of the FUV data is AIS. We retain these galaxies for the study, because
the regions over which we are calculating \ssfr\  for this sample correspond to 
$\sim$100 kpc$^{\rm -2}$ ($\sim$20 arcsec beams), so the signal-to-noise for 
the total FUV flux in each such region is large enough even for AIS data to be usable.

Some properties of the `Bluedisk' sample are listed in Table~\ref{tab:sampB}.
The upper section of the table lists the 'Bluedisk' main sample, while   
the lower section of the table lists the `control' sample.
Column (1) gives the galaxy identifier, (2) the derived stellar mass, (3) the derived HI mass, (4) the distance to the galaxy in Mpc, (5) the measured axial ratio in the $r$ band \citep[used when calculating the surface densities as described in Section~\ref{ssec:est} and given here due to changes compared to values tabulated in][]{wan13}, and (6) the size of the HI beam in kpc.
All values are taken from \citet{wan13}.

\begin{table}
\caption{Bluedisk spiral galaxy sample}
\label{tab:sampB}
\begin{tabular}{|cccccc|}
\hline
ID&M$_{*}$&M$_{\rm HI}$&D&b/a&HI beam\\
  &(10$^{\rm 10}$~M$_{\odot}$)&(10$^{\rm 9}$~M$_{\odot}$)&(Mpc)&&(kpc$\times$kpc)\\
\hline
Main\\
1 &2.69&12.30&124.23&0.64&16.2$\times$10.2\\
2 &4.17& 8.71&110.49&0.41&13.9$\times$ 9.2\\
3 &2.63& 7.76&128.88&0.86&16.1$\times$10.8\\
4 &3.72&18.20&115.06&0.39&17.5$\times$ 9.1\\
5 &2.09&15.85&110.56&0.89&14.6$\times$ 9.4\\
6 &6.92&20.42&110.55&0.52&15.1$\times$ 9.1\\
8 &1.91&15.85&119.64&0.74&13.8$\times$10.3\\
12&4.27&13.18&115.05&0.45&10.6$\times$ 9.1\\
14&6.17&13.80&105.97&0.49&11.3$\times$ 9.0\\
15&6.61&24.55& 96.85&0.73&13.7$\times$ 7.9\\
16&2.04&10.00&119.70&0.68&12.6$\times$10.2\\
17&4.90&21.88&124.21&0.68&12.8$\times$11.1\\
18&2.19&11.75&101.46&0.45&12.5$\times$ 9.8\\
19&1.78&14.79&115.06&0.81&14.2$\times$ 8.9\\
20&1.62&16.22&115.12&0.53&14.1$\times$ 9.0\\
21&2.57& 7.59&133.45&0.60&17.3$\times$10.7\\
22&6.61&10.00&124.31&0.28&16.0$\times$ 9.6\\
24&5.13&16.22&133.42&0.48&13.0$\times$11.4\\
26&2.04& 3.16&101.45&0.39&12.4$\times$ 8.5\\
30&2.45&11.75&124.27&0.26&16.0$\times$10.0\\
35&4.07& 8.51&106.01&0.36&12.3$\times$ 9.1\\
47&3.39&10.23&105.99&0.60&15.3$\times$ 7.9\\
50&7.41& 6.31&124.24&0.29&11.8$\times$10.2\\
\hline     
Control\\                                                                  
9 &5.75& 8.13&124.21&0.45&13.5$\times$ 9.1\\
10&8.13&10.47&128.91&0.89&12.8$\times$11.1\\
11&4.37& 5.13&105.99&0.61&11.2$\times$ 8.4\\
23&5.62& 8.71&133.41&0.43&15.2$\times$ 9.7\\
25&3.47& 5.75&124.21&0.78&12.0$\times$10.2\\
27&2.00& 1.86&128.84&0.75&17.8$\times$10.0\\
28&3.47& 1.32& 83.33&0.86& 9.0$\times$ 6.9\\
32&1.95& 3.63&119.67&0.44&17.0$\times$ 9.4\\
33&5.50& 1.58&119.67&0.57&13.1$\times$10.4\\
36&2.19& 2.45&110.53&0.65&11.3$\times$ 9.4\\
37&2.45& 5.01&115.10&0.69&11.5$\times$ 9.8\\
38&5.62& 0.56&110.48&0.73&10.9$\times$ 9.6\\
39&6.92& 1.38&105.98&0.75&13.9$\times$ 8.3\\
40&2.29& 4.47&119.70&0.84&17.1$\times$ 9.9\\
42&6.31& 3.72&124.28&0.57&13.6$\times$10.7\\
43&1.82& 3.89&119.67&0.73&15.0$\times$ 9.9\\
44&6.03& 1.91&124.25&0.64&14.5$\times$ 9.9\\
45&4.79& 4.07&115.09&0.39&11.9$\times$ 9.6\\
49&2.63& 5.50&133.42&0.62&20.6$\times$10.9\\
\hline
\end{tabular}
\end{table}

\section {Analysis}
\label{sec:ana}

\subsection {Overview}
\label{ssec:ovv}

We derive relations between \ssfr\ and  \sgas\ or \shi\ in the HI-dominated regime for the galaxies in our three samples.  
Here \shi\ refers to the atomic gas surface density corrected for the presence of helium by multiplying the HI surface density by a factor of 1.34, and \sgas\ refers to the total gas surface density including the contribution from H$_{\rm 2}$.

We begin with (i) THINGS spiral galaxies at the resolution of the CO maps (mean and median resolution of $\sim$600 pc)
by averaging over 1 kpc sized regions.
Next, we decrease the size of the regions over which the surface densities are compared.
We consider (ii) THINGS spiral galaxies at a resolution of $\sim$400 pc and,  (
iii) FIGGS dwarf irregular galaxies at a resolution of $\sim$400 pc.
400 pc is a natural and convenient choice of resolution for the $L_*$ spiral sample,
because 7 of the 12 THINGS  galaxies are  observed in HI at a 
comparable native resolution.
The resolution of the {\it Spitzer} 24 \um\ data is $\sim$6$^{\prime\prime}$, i.e. comparable to the HI.
In the THINGS spiral sample, we choose HI-dominated regions by referring to the CO maps and checking whether there is any detectable CO emission.
This procedure ensures that not only do we choose choose regions from the outer disks of the galaxies, but also from the inter-arm regions of these spiral galaxies. 
We consider all regions in dwarfs, including the very central regions, as being dominated by HI.

Finally we study the relation between \ssfr\ and \sgas\  at large ($\sim$10 kpc) scales
using the  `Bluedisk' main and control spiral galaxies at the resolution of their HI maps.
In this sample, all regions whose centres are within a distance of R25 
from the optical centre of the galaxy  are not considered. Once again, this is done
in order to restrict the analysis to regions where the HI gas is likely to dominate.
For all datasets, fluxes are extracted from matched square grid cells in the gas maps and the UV/IR images.
Adjacent regions overlap by half the length of each side in order to ensure Nyquist sampling.

\subsection{Measuring surface densities of gas}
\label{ssec:gas}

For the THINGS $/$ HERACLES galaxies the flux in CO maps is converted to \shm\ using the Galactic CO-to-H$_{\rm 2}$ conversion factor of 4.35 ${\rm M_{\odot}~pc^{-2}~(K~km~s^{-1})^{-1}}$.
The limiting \sgas\ for HERACLES CO data is $\sim$3 ${\rm M_{\odot}~pc^{-2}}$ \citep{ler09}.
Only \sgas\ values above this limit are considered.

HI maps are smoothed to the resolution of CO maps for the 1 kpc scale study.
For all other datasets HI maps are used at their native resolutions.
The flux density in HI is converted to column density using the standard transformation for 
emission from an optically thin medium.
The limiting \shi\ for THINGS galaxies is $\sim$0.5 ${\rm M_{\odot}~pc^{-2}}$ \citep{wal08}, 
and only regions with \shi\ greater than this value are considered.
The 3$\sigma$ HI column density sensitivity threshold for FIGGS galaxies varies from 0.4 to 5 ${\rm M_{\odot}~pc^{-2}}$ with a median of 1.4 ${\rm M_{\odot}~pc^{-2}}$, and only regions in FIGGS galaxies with \shi\ greater than the corresponding threshold are used in our study.
The 3$\sigma$ HI column density sensitivity thresholds corresponding to the 2$\sigma$ values 
given in Column (7) of Table 2 in \citet{wan13} are used as cut-off for `Bluedisk' sample galaxies and only regions with higher \shi\ are considered. 

\subsection{Measuring surface densities of SFR}
\label{ssec:est}

\citet{daS14} have shown that \ha\ becomes a highly unreliable tracer of SFR at low SFR.
Therefore, we use FUV emission from {\it GALEX} (resolution $\sim$4.5$^{\prime\prime}$), and use mid--infrared 24 \um\ emission from {\it Spitzer} (resolution $\sim$6$^{\prime\prime}$) or 22 \um\ emission from {\it WISE} (resolution $\sim$12$^{\prime\prime}$) to correct our SFRs for attenuation due to dust.
Foreground stars and
background galaxies are identified and masked in the
FUV and 24 \um\ maps of THINGS and FIGGS galaxies. 
For the THINGS galaxies, this is done by applying the colour based masks of \citet{mun09} followed by visual inspection.
For the FIGGS galaxies, the masking is done by visual inspection.
The masking of the {\it WISE} 22 \um\ images of the `Bluedisk' sample galaxies is done
as described in \citet{hua14}.
The PSF of {\it Spitzer} 24 \um\ maps, and to a lesser extent that of the {\it GALEX} FUV maps have extended sidelobes.
We use the kernels provided in \citet{ani11} to convert to Gaussian beams 
close to the final required beam (that of the CO or HI map) for the 24 \um\ and FUV maps,
 and in the next step we smooth the Gaussian beam to match that of the CO or HI map.
The {\it WISE} 22 \um\ maps are directly smoothed to the resolution of the corresponding HI map.
Fluxes are then extracted from regions of requisite size in these smoothed maps.

The conversion between FUV luminosity and SFR is given by \citet{hao11}:
\begin{equation}
\rm {log~\frac{SFR}{M_{\odot}~yr^{-1}}~=~log~\frac{\nu}{Hz} \frac{L_{\nu}}{ergs~s^{-1}~Hz^{-1}}~-~43.35}
\label{eqn:fuv}
\end{equation}
\noindent 
We account for the energy from star formation re-radiated at infra-red wavelengths due to dust using mid--infrared data \citet{hao11}:
\begin{equation}
\rm {L_{FUV,corr}~=~L_{FUV, obs}~+~3.89~L_{25~\mu m}}
\label{eqn:24um}
\end{equation}
\noindent which is fed into equation~\ref{eqn:fuv} to obtain the SFR.
The calibration is valid for either Salpeter or Kroupa IMFs with stellar mases in the range of 0.1--100 M$_{\odot}$ and solar metallicity.
Only fluxes 3 times higher than the corresponding noise levels for {\it Spitzer} 24 \um\ and {\it WISE} 22 \um\ data are considered, as they are only additive corrections to the FUV flux.
The above calibration is also applicable to dwarf galaxies as discussed in \citet{DeL14}.

For FIGGS sample galaxies with 24 \um\ detections, we compare the detectable infrared fluxes in 400 pc regions to the FUV luminosity in the same regions.
A bivariate regression fit to the data is then used to estimate the 24 \um\ flux in a region of a galaxy without {\it Spitzer} 24 \um\ measurements:
\begin{equation}
\rm{\log \frac{F_{24 \mu m}}{MJy~st^{-1}}~=~1.67 \log \frac{L_{FUV}}{ergs~s^{-1}~Hz^{-1}}~-~39.8}
\label{eqn:irf}
\end{equation}
\noindent This estimate is only derived for regions with sufficiently high FUV luminosity (${\rm >~24~ergs~s^{-1}~Hz^{-1}}$).

To estimate the error on the measured SFRs we add the following terms in quadrature: measurement error for FUV and 24 \um\ fluxes (as determined from the `relative response' and `weight' maps respectively), 10\% flux calibration error for {\it GALEX} FUV data, 5\% flux calibration error for {\it Spitzer} 24 \um\ data, and 50\% error to account for the uncertainty in the SFR calibration caused by variations in the IMF and star formation history following \citet{ler12,ler13}.

The measured FUV flux used in equation~\ref{eqn:24um} also requires a correction if star formation is happening at very low metallicities as is the case for the FIGGS sample dwarfs (viz. Z~$\sim 0.1$~Z$_{\odot}$~as can be seen from Table~\ref{tab:sampF}).
\citet{rai10} used evolutionary synthesis models using a Salpeter IMF and constant star formation for the last 10$^{\rm 8}$ years to estimate emergent fluxes in sub-solar metallicity environments and found that FUV fluxes increase by $\sim$ 11\%, 19\%, 27\%, 32\% for metallicities of  0.4, 0.2, 0.05, 0.02 times solar.
For each FIGGS sample galaxy we do a linear interpolation between the \citet{rai10} values stated above and arrive at the percentage increase and hence the correction factor for the emergent FUV flux at the metallicity of the galaxy.
For spiral galaxies the metallicity at the very outskirts of the disc reduces at the most to 0.4-0.6 Z$_{\odot}$ according to available evidence from observations within our Galaxy \citep{car07,gen14}.
This also seems to be the case for the THINGS sample galaxies \citep{mou10} and the `Bluedisk' sample galaxies \citep{car15} with available metallicity gradient measurements.
For this metallicity the correction factor is only $\sim$10\%.
This is small compared to the total error on the SFR calibration, and therefore we do not apply any correction for the reduction in ISM metallicity in the outskirts of the spiral galaxies in our sample.

\section{Results and Discussion}
\label{sec:res}

\subsection{Deriving the K-S relations for the different datasets}

In each logarithmic bin of \sgas\ / \shi\ we determine and plot: 
(i) the logarithms of the median ({\it median}~(Log~$\Sigma_{\rm SFR}$)), 5 and 95 percentile \ssfr s taking into consideration all regions including those with zero or negative fluxes, (ii) the logarithm of the average \ssfr\ calculated by summing over the fluxes from all regions including those with zero or negative fluxes: Log~({\it average}~$\Sigma_{\rm SFR}$).
\citet{daS14} show that due to the reasons discussed in Section~\ref{sec:int}, if the SFR being traced is low, using standard multiplicative calibrations to convert a measured FUV flux to SFR will give erroneous results.
Even though the median flux characterizes the distribution of measured fluxes in a robust manner, if the median flux originates from a low SFR region the {\it median}~(Log~$\Sigma_{\rm SFR}$) determination is likely to be biased to a lower value than the true underlying SFR. In the HI dominated regime under investigation, most fluxes from individual regions are in this uncertain regime.
A comprehensive approach of quantifying the relation between surface densities of gas and star formation in this regime will involve correcting for the bias in the measured SFR using the output from simulations like the one by \citet{daS14}.
Their simulations show that although the bias in measured SFR is generic, the exact amount of the bias depends critically on quantities like the total fraction of stars forming in clusters.
Such quantities are as yet not well constrained from observations for dwarf galaxies and in the outskirts of spirals, and therefore it is beyond the scope of this paper to determine the corrected statistical distribution of $\Sigma_{\rm SFR}$~in the low SFR regime.
In this study we therefore use the alternative determination Log~({\it average}~$\Sigma_{\rm SFR}$), for which the summed total SFR in the bin always remains above $\rm{log(SFR/(M_{\odot}~yr^{-1}))}~\sim~-$ 3, where FUV should be a highly reliable SFR estimator \citep{daS14}.
We note that we have also checked that the distribution of negative flux values is well approximated by a one-sided Gaussian function, and that there are no significant tails at large negative values that compromise the summation.

We carry out similar investigations for our different datasets covering different galaxy samples and scales.
For each dataset merely the Log~({\it average}~$\Sigma_{\rm SFR}$) values are considered for determining Kennicutt-Schmidt relations.
In each case the relation is determined using a linear regression fit to the Log~({\it average}~$\Sigma_{\rm SFR}$) and binned Log~(\shi) values, only considering bins with a statistically significant number of individual regions in them ($\sim$ 100 or more).
It is to be noted that a relation thus determined is representative of the `average' trend, and the error on the fit is only indicative of the variation in any such `average' relation and is not indicative of the scatter that might be present in the \ssfr s from low \shi\ regions.
Also the Kennicutt-Schmidt relations determined here are subject to change depending on correction of low \ssfr\ data for bias as discussed above, and statistically robust modelling to fit the corrected dataset after taking into consideration the full scatter in measured \ssfr s.
A distinctive feature of our study is that we select and study the Kennicutt-Schmidt relation for the {\it entire} HI-dominated ISM of the galaxy (discussed below), as opposed to concentrating on regions in the disk within some fixed annuli.

\subsubsection*{1 kpc scale}

\begin{figure}
\begin{center}
\psfig{file=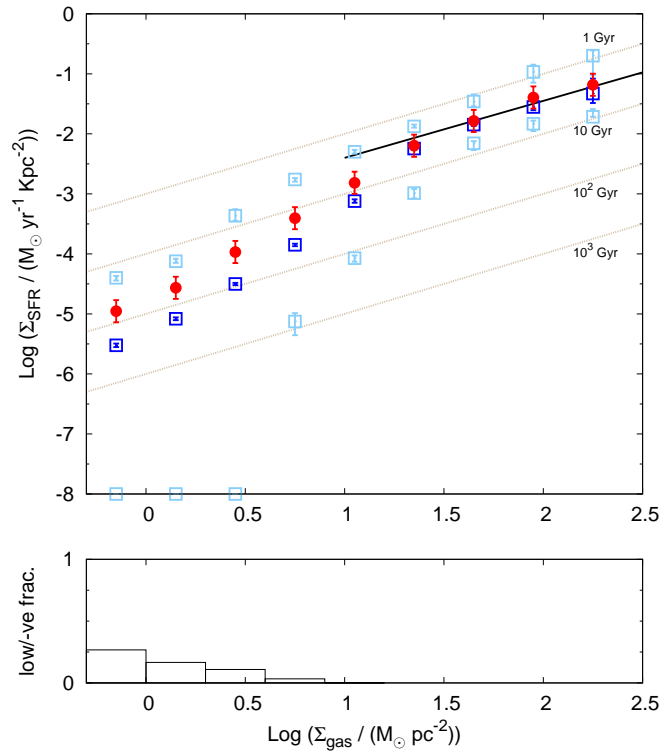,width=3.5truein}
\end{center}
\caption{The logarithms of the surface densities of SFR and gas (atomic+molecular) plotted against each other for 1 kpc sized regions in THINGS sample galaxies. The open dark blue squares (in the upper panel) represent the {\it median}~(Log~$\Sigma_{\rm SFR}$) (see text) in each 0.3 dex \sgas\ bin and the open light blue squares represent the 95 and 5 percentile points (shown at y=$-$8 if negative), with errorbars determined by bootstrapping the data in each \sgas\ bin. In the upper panel filled red circles represent the Log~({\it average}~$\Sigma_{\rm SFR}$) (see text) with associated errorbar. In the upper panel the bold line at the high \sgas\ end is the \shm\ based K--S law for nearby spirals from \citet{ler13}. Dotted beige lines in the background indicate various constant gas depletion timescales. The lower panel shows the fraction of points lying below the plotted range including those with negative FUV flux in each \sgas\ bin.}
\label{fig:t1kpctot}
\end{figure}

We first show results for THINGS spiral galaxies over 1 kpc sized regions in Figure~\ref{fig:t1kpctot}.
For this dataset we have measurements of molecular gas, and can do a direct comparison of our results with previous results which used the same sample.
At the high \sgas\ end where molecular hydrogen dominates and the SFRs in individual regions are high, our determinations of both {\it median}~(Log~$\Sigma_{\rm SFR}$) and Log~({\it average}~$\Sigma_{\rm SFR}$) are consistent with each other and our results match those of \citep{ler13} very well.
At low valued of \sgas\,  the scatter in values measured in individual regions increases, 
and the number of regions with very low or negative FUV flux and below detection limit increases.
We find that the K--S relation for the outer discs of spirals derived  by \citet{big10} matches our determination of the {\it median}~(Log~$\Sigma_{\rm SFR}$).
For low \sgas\ bins, however,  the {\it median}~(Log~$\Sigma_{\rm SFR}$) diverges from the Log~({\it average}~$\Sigma_{\rm SFR}$) (middle panel of Figure~\ref{fig:t1kpctot}). 
The difference between the Log~({\it average}~$\Sigma_{\rm SFR}$) and {\it median}~(Log~$\Sigma_{\rm SFR}$) reaches almost an order-of-magnitude
at the lowest gas surface densities.
The Log~({\it average}~$\Sigma_{\rm SFR}$) does not exhibit as steep a decline as the {\it median}~(Log~$\Sigma_{\rm SFR}$), and the average gas consumption 
timescale remains around few tens of Gyrs at the lowest gas column densities we can trace (1 $M_{\odot}$ pc$^{-2}$).

\begin{figure}
\begin{center}
\psfig{file=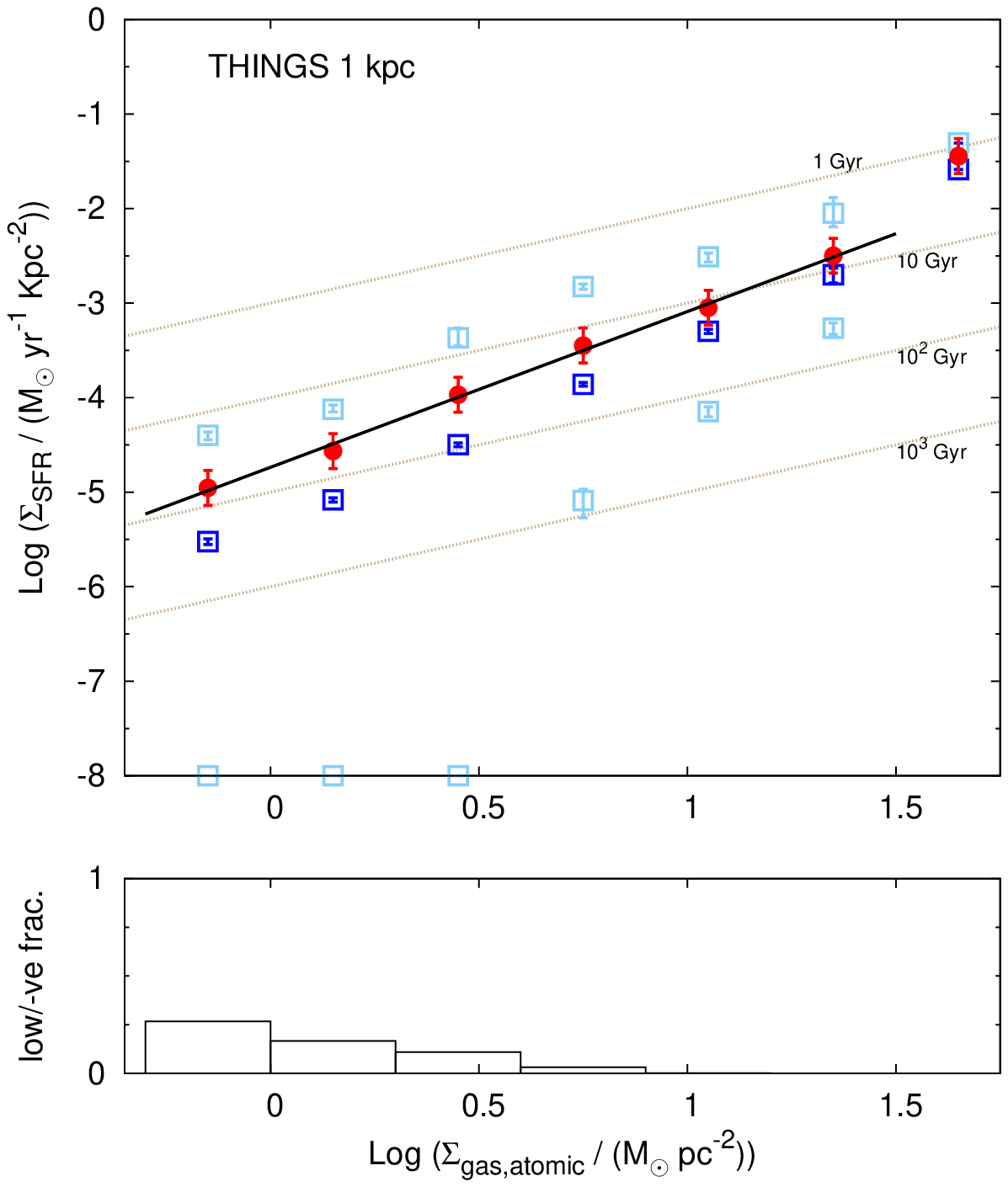,width=3.5truein}
\end{center}
\caption{The logarithms of the surface densities of SFR and atomic gas plotted against each other for 1 kpc sized regions in THINGS sample galaxies. The bold black line represents the best fit Kennicutt-Schmidt relation (see text). Other symbols are similar to those in Figure~\ref{fig:t1kpctot}.}
\label{fig:t1kpc}
\end{figure}

The results for the HI dominated regions at 1 kpc scales in THINGS spirals are shown in Figure~\ref{fig:t1kpc}, and the best fit values to the binned Log~({\it average}~$\Sigma_{\rm SFR}$) -- Log~($\Sigma_{\rm gas,atomic}$) values is listed in Table~\ref{tab:ncomp}.
We have repeated the study by smoothing all data to exactly 1 kpc resolution, and found differences in the results to be insignificant.

\subsubsection*{400 pc scale}

\begin{figure}
\begin{center}
\begin{tabular}{c}
{\mbox{\includegraphics[width=3.5truein]{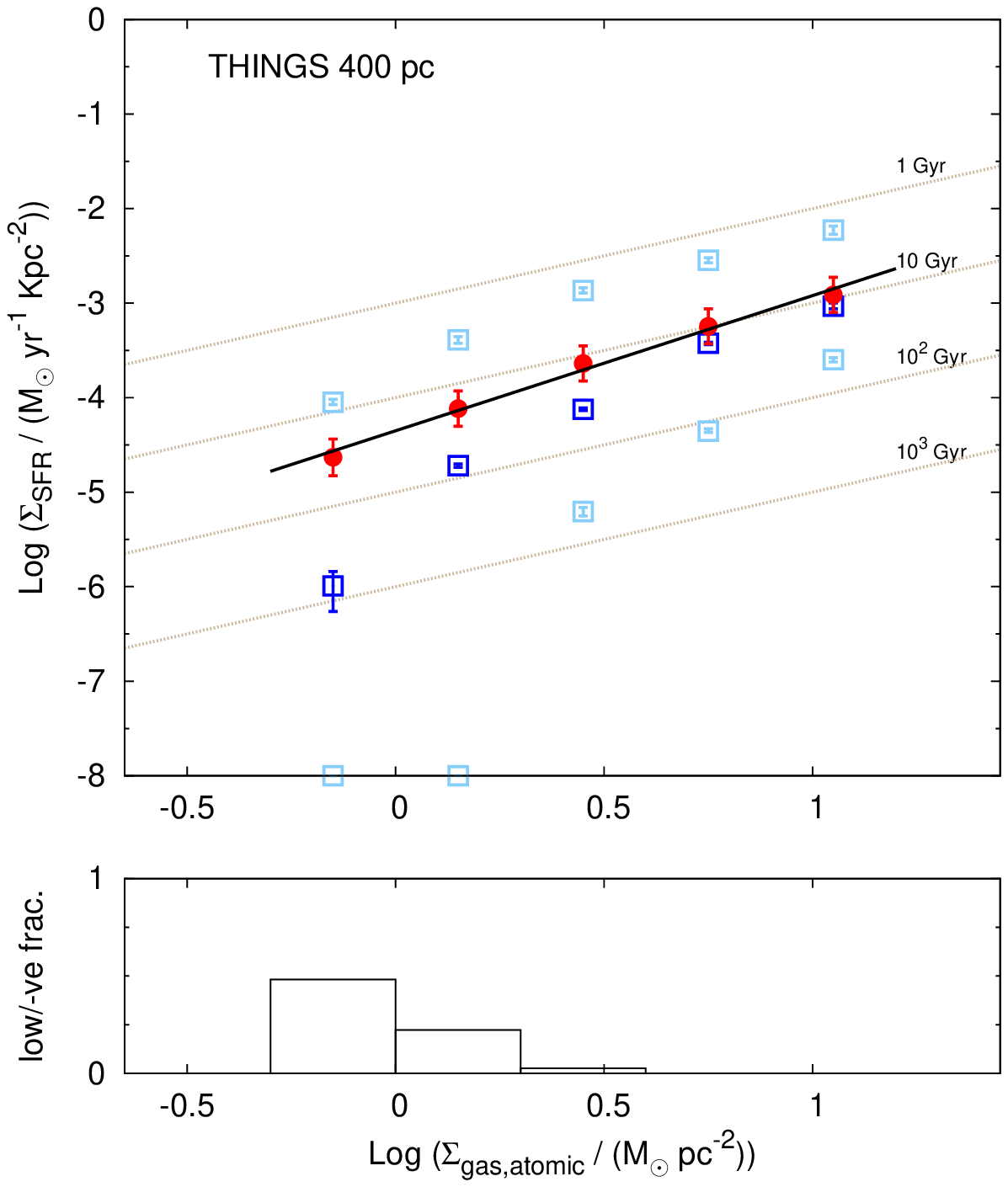}}}\\
{\mbox{\includegraphics[width=3.5truein]{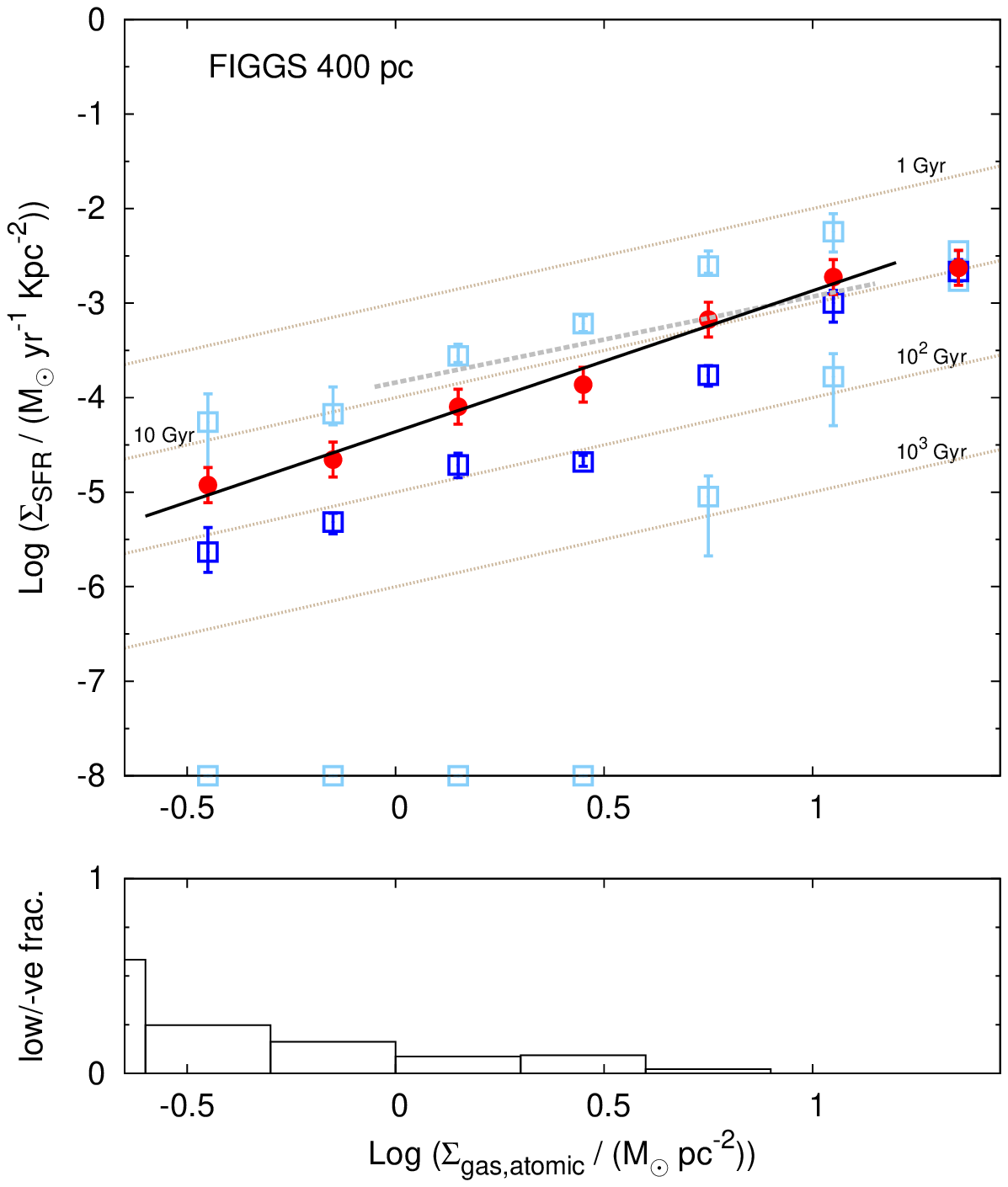}}}\\
\end{tabular}
\caption{The logarithms of the surface densities of SFR and atomic gas plotted against each other for 400 pc sized regions in THINGS (upper half) and FIGGS (lower half) sample galaxies. The bold black lines represent the best fit Kennicutt-Schmidt relations (see text). The grey dashed line in the FIGGS sample panel is the best fit relation for disk-averaged values from \citet{roy14}. Other symbols are similar to those in Figure~\ref{fig:t1kpctot}.}
\end{center}
\label{fig:400pc}
\end{figure}

The results at 400 pc scales in THINGS spirals and FIGGS dwarfs are shown in Figure~\ref{fig:400pc} and the best fit values tabulated in Table~\ref{tab:ncomp}.
An interesting point to note is that for the lowest \shi\ bin in the THINGS sample plot, a large fraction of regions with low or negative \ssfr s results in the {\it median}~(Log~$\Sigma_{\rm SFR}$) to decrease sharply compared to the value in the immediately higher \shi\ bin, whereas the decrease in Log~({\it average}~$\Sigma_{\rm SFR}$) compared to the value in the higher \shi\ bin is moderate and follows the determined Kennicutt-Schmidt relation.
As can be seen from the figure, we restrict the fit to \shi\ bins with small fraction of regions with noisy \ssfr.
The apparent mismatch between the disc-averaged K--S relation seen in FIGGS galaxies \citep{roy14} 
and the steeper relation we find here at 400 pc scales is explainable 
by the fact that for the disc-averaged study we bias ourselves
to $\sim$kpc sized regions which are forming stars (the optical disc).
The extended HI with little or no star formation present in most of the FIGGS galaxies was excluded in the \citet{roy14} study.

\subsubsection*{10 kpc scale}

\begin{figure}
\begin{center}
\begin{tabular}{c}
{\mbox{\includegraphics[width=3.5truein]{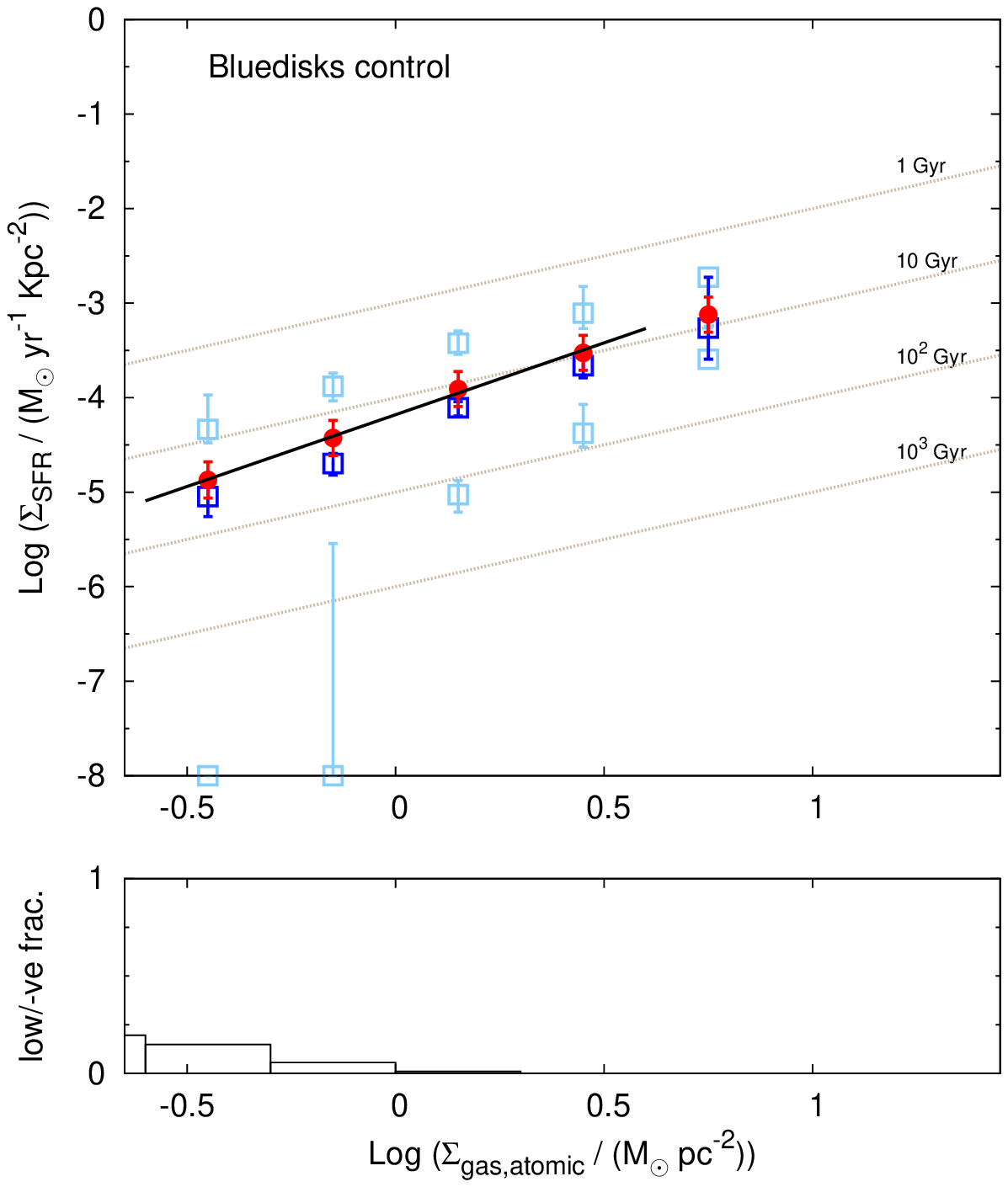}}}\\
{\mbox{\includegraphics[width=3.5truein]{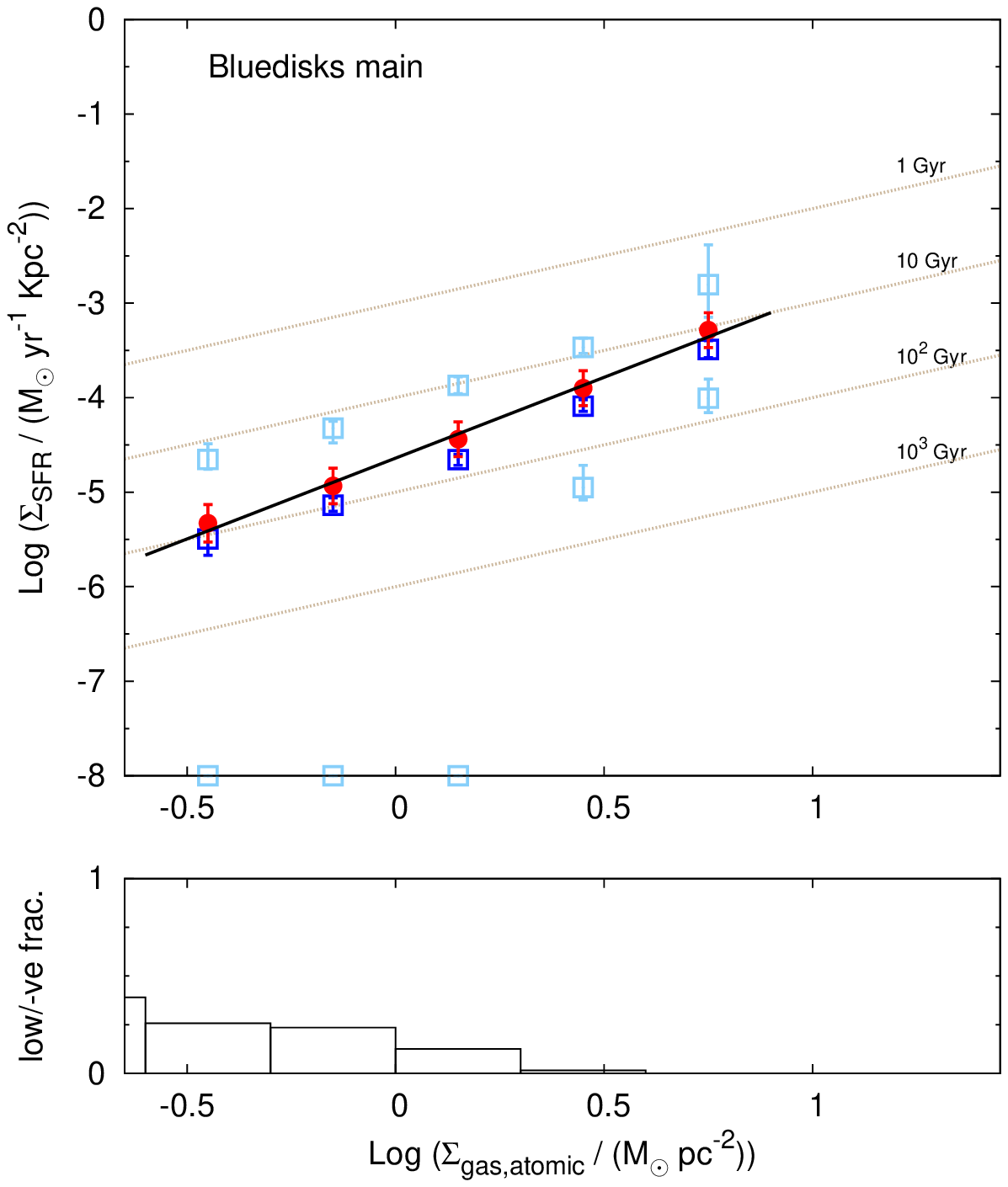}}}\\
\end{tabular}
\caption{The logarithms of the surface densities of SFR and atomic gas plotted against each other for HI beam sized regions in `Bluedisk' control (upper half) and main (lower half) sample galaxies. The bold black lines represent the best fit Kennicutt-Schmidt relations (see text). Other symbols are similar to those in Figure~\ref{fig:t1kpctot}.}
\end{center}
\label{fig:bd}
\end{figure}

The results for the `Bluedisk' sample spiral galaxies are shown in Figure~\ref{fig:bd} and the best fit values tabulated in Table~\ref{tab:ncomp}.
The differences in the K--S relation seen in `Bluedisk' main and control galaxies is worth noticing.
For the main sample the large amounts of HI outside the optical disc 
which have negligible or no star formation going on in them appears to steepen the relation at
low gas surface density.

In order to compare the results from the `Bluedisk' spirals with that obtained for the nearby THINGS galaxies, we tried smoothing the data for THINGS galaxies to 10 kpc scales. But this did not result in a trustworthy result, as the FUV maps had to be smoothed 20 times or more which resulted in the final FUV maps to be dominated by noise except in the very central regions of the sample galaxies.
We are also aware of the fact that if the THINGS galaxies were to be imaged at 10 kpc resolutions, some low level highly extended emission may be present which is not picked up at their present $\sim$ kpc resolution.
Smoothing the data does not alleviate this issue.
Notwithstanding the existence of these sources of uncertainty, a linear fit to the final binned data for THINGS galaxies smoothed to 10 kpc resolution matched the values derived for the `Bluedisk' control sample well.

\begin{table}
\begin{center}
\caption{Comparing derived Kennicutt-Schmidt relations}
\label{tab:ncomp}
\begin{tabular}{|lcc|}
\hline
Dataset&slope&intercept\\
\hline
FIGGS 400pc     &1.5~$\pm$0.1~&$-$4.36$\pm$0.06\\
THINGS 400 pc   &1.43$\pm$0.07&$-$4.35$\pm$0.05\\
THINGS 1 kpc    &1.65$\pm$0.04&$-$4.74$\pm$0.03\\
Bluedisk control&1.52$\pm$0.06&$-$4.18$\pm$0.02\\
Bluedisk main   &1.71$\pm$0.08&$-$4.64$\pm$0.04\\
\hline
\end{tabular}
\end{center}
\end{table}

\subsubsection*{Quantifying the scatter}

\begin{table}\centering
\caption{Statistics for \ssfr\ -- \shi\ plane}
\label{tab:stat}
\begin{tabular}{@{}lccccc@{}}\toprule
Log $\Sigma_{\rm gas, atomic}$&diff.&\multicolumn{4}{c}{percentile $\Sigma_{\rm SFR}$~in $M_{\odot}~yr^{-1}~kpc^{-2}$}\\
\cmidrule{3-6} 
($M_{\odot}~pc^{-2}$)&(dex)&95&84&16&05\\ \midrule
FIGGS 400 pc\\
$-$0.45&0.71&6e-05&8e-06&$-$8e-07&$-$1e-05\\
$-$0.15&0.66&7e-05&9e-06&$-$4e-07&$-$3e-06\\
~~~0.15&0.62&3e-04&8e-05&~~~2e-06&$-$1e-06\\
~~~0.45&0.82&6e-04&2e-04&~~~3e-06&$-$2e-06\\
~~~0.75&0.59&2e-03&6e-04&~~~3e-05&~~~9e-06\\
~~~1.05&0.27&6e-03&2e-03&~~~3e-04&~~~2e-04\\
\hline
THINGS 400 pc\\
$-$0.15&undef.&9e-05&2e-05&~~~~~0&$-$1e-05\\
~~~0.15&0.60&4e-04&9e-05&~~~~~0  &$-$8e-06\\
~~~0.45&0.49&1e-03&4e-04&~~~2e-05&~~~6e-06\\
~~~0.75&0.18&3e-03&1e-03&~~~1e-04&~~~4e-05\\
~~~1.05&0.12&6e-03&3e-03&~~~4e-04&~~~3e-04\\
\hline
THINGS 1 kpc\\
$-$0.15&0.57&4e-05&8e-06&$-$1e-06&$-$6e-06\\
~~~0.15&0.52&8e-05&3e-05&~~~1e-07&$-$7e-06\\
~~~0.45&0.53&4e-04&1e-04&~~~5e-06&$-$8e-06\\
~~~0.75&0.41&1e-03&6e-04&~~~4e-05&~~~8e-06\\
~~~1.05&0.25&3e-03&1e-03&~~~2e-04&~~~7e-05\\
~~~1.35&0.20&9e-03&5e-03&~~~8e-04&~~~5e-04\\
\hline
Bluedisk control\\
$-$0.45&0.18&5e-05&3e-05&~~~1e-06&$-$2e-05\\
$-$0.15&0.27&1e-04&7e-05&~~~7e-06&$-$8e-07\\
~~~0.15&0.20&4e-04&2e-04&~~~2e-05&~~~9e-06\\ 
~~~0.45&0.13&8e-04&5e-04&~~~9e-05&~~~4e-05\\
\hline
Bluedisk main\\
$-$0.45&0.17&2e-05&1e-05&$-$2e-06&$-$1e-05\\
$-$0.15&0.20&5e-05&2e-05&$-$3e-06&$-$1e-05\\
~~~0.15&0.22&1e-04&6e-05&$-$4e-06&$-$9e-06\\
~~~0.45&0.19&3e-04&2e-04&~~~3e-05&~~~1e-05\\
~~~0.75&0.20&2e-03&7e-04&~~~2e-04&~~~1e-04\\
\bottomrule
\end{tabular}
\end{table}

In Table~\ref{tab:stat} we show how the offset between measured \ssfr s and their scatter in the \ssfr\ -- \shi\ plane varies with \shi\ and between datasets.
Negative \ssfr\ values are arrived at using the SFR calibration on negative fluxes and have no physical meaning, but are used here to be indicative of the scatter.
Column (1) gives the centre of the \shi\ bin in log, (2) the difference in dex between the Log~({\it average}~$\Sigma_{\rm SFR}$) and {\it median}~(Log~$\Sigma_{\rm SFR}$) values.  
For `Bluedisk' galaxies where fluxes are averaged over $\sim$10 kpc sized regions (equivalent to summing over fluxes from hundreds of 400 pc or 1 kpc sized regions), the difference is the least and match well considering the errors on the measured Log~({\it average}~$\Sigma_{\rm SFR}$)s. 
For datasets with 400 pc and 1 kpc sized regions the difference increases with decreasing \shi, and the offset is more at the same \shi\ bin for smaller sized regions.
Columns (3) through (6) list the 95, 84, 16, and 05 th percentile \ssfr\ value measured in different regions within a respective \shi\ bin.
The scatter in measured \ssfr s obviously increases with decreasing \shi, and decreasing region size.
But there is no clear indication that the scatter increases for dwarf galaxies compared to spirals for the same region size.

\subsection{A universal K--S relation in HI dominated ISM}

\begin{figure}
\begin{center}
\psfig{file=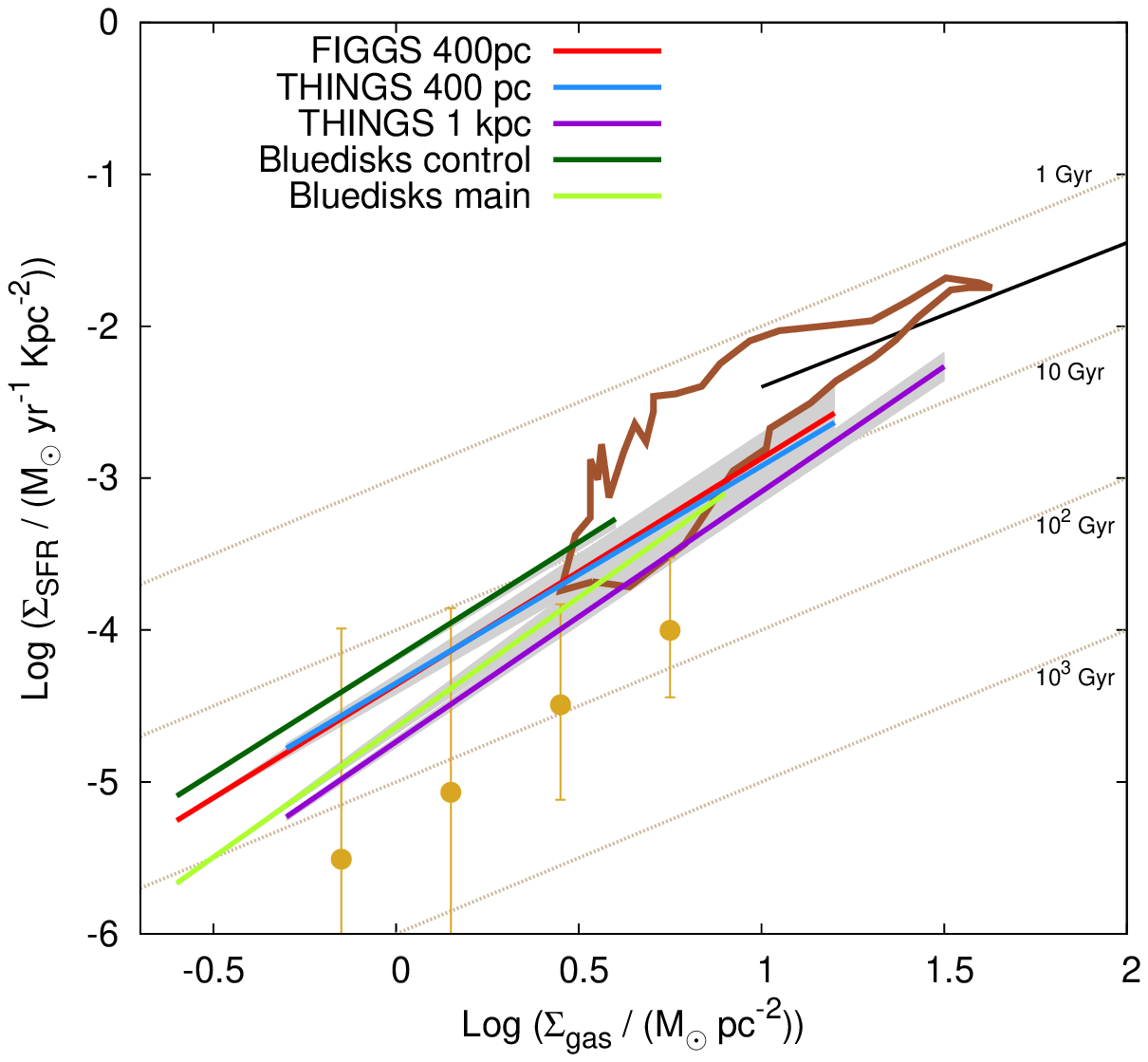,width=3.5truein}
\end{center}
\caption{The resolved Kennicutt-Schmidt relation in the HI dominated regime, in relation to other determinations. The black line represents the \shm\ based K--S law for nearby spirals from \citet{ler13}, the brown contour indicates the relation as seen on 750 pc scales within the optical discs of nearby spirals by \citet{big08} (the red shaded area in the middle right panel of their Figure 8), and the yellow open circles represent the median \ssfr\ in similar \sgas\ bins in the very outer discs of spirals from the THINGS survey measured by \citet{big10} at 1 kpc scales with the scatter indicated by the errorbars. The K--S type relations for the different datasets studied in these paper are marked by different colours and shown in the range of \shi\ within which the respective linear fits were done. The 1 $\sigma$~error on the fits and their overlap are represented by the grey shaded area. Dotted beige lines in the background indicate various constant gas depletion timescales.}
\label{fig:HIks}
\end{figure}

In Figure~\ref{fig:HIks} our fits to the Kennicutt-Schmidt relation in the HI dominated ISM of various types of galaxies and at different resolutions, are compared to previous determinations of the relation between \ssfr\ and \sgas\ in spirals.
The linear fit to the relation seen 
for outer discs of spirals by \citet{big10} is also provided as reference, although in their study only regions within an annulus with inner radius of R25 and outer radius of 2 times R25 were considered. 
From the values listed in  Table~\ref{tab:ncomp}, we can see that a K--S type relation in HI dominated gas with a power law 
slope of $\sim$1.5 is a good fit to the data for both $\sim L_*$ spirals and
dwarf galaxies. 
The derived K--S relation is always much more inefficient than that between molecular/total gas and 
SFR within the optical disks of $L_*$ spirals. The amplitude of the relations for different samples agree within a factor of $\sim 2$.

\begin{figure}
\begin{center}
\psfig{file=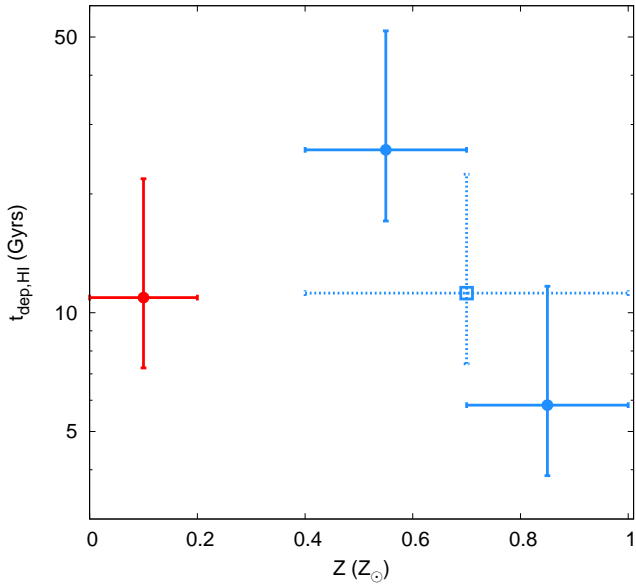,width=3.5truein}
\end{center}
\caption{The variation of HI depletion time with metallicity when measured at 400 pc resolution. The red point and errorbars represent the value for the FIGGS sample. The blue filled points and bold errorbars represent the values for THINGS galaxies (400 pc resolution) in two different metallicity ranges. The average value for all HI-dominated regions in the THINGS 400 pc dataset is represented by the blue open square with dashed errorbars.}
\label{fig:met}
\end{figure}

We have derived the Kennicutt-Schmidt relation in the HI dominated ISM of three very different environments -- dwarf galaxies, outer disks of normal
galaxies and outer disks of `Bluedisk' galaxies -- and found them to remarkably similar.
One of the properties which characterizes the diversity of the ISM in the different datasets studied, is their metallicity.
In Figure~\ref{fig:met} we check for any variation of HI depletion time with metallicity.
For clarity we show the results for two datasets mapped at the same resolution: FIGGS galaxies and THINGS galaxies mapped at 400 pc resolution.
We do not have comprehensive measurements of the metallicities for all the regions in our sample galaxies, but deduce the expected ranges in metallicity sampled by the different datasets using existing observations and trends. 
For the FIGGS sample galaxies the metallicities are assumed to span the range of values tabulated in Table~\ref{tab:sampF}, mindful of the fact that dwarf galaxies do not show appreciable metallicity gradients \citep{wes13}.
For the THINGS sample galaxies the lower limit in metallicity is defined by the observed flattening in the metallicity gradient beyond R25 in nearby spirals \citep{ber13,bre12} and in our Galaxy \citep{gen14}.
The upper range in metallicity for THINGS galaxies is determined by the average measured metallicity at the maximum radius from the centre of the sample galaxies where CO emission in observed, for the galaxies with existing metallicity gradient measurements \citet{mou10}. 
We use the average of the metallicities (in solar metallicity units) derived using the two different metallicity gradients \citep[following KK04 and PT05 methods, see][for details]{mou10}. 
There are some HI-dominated regions at radii smaller than the maximum extent of CO in each sample galaxy disk, but these regions are inter-arm regions. 
The metallicity in inter-arm regions of spiral galaxies are lower than the metallicity measured from HII regions in the spiral arms \citep[e.g][]{ced12}, and we therefore assume that the metallicity of these interarm regions are at the most solar.
The HI-dominated regions from the THINGS dataset are divided into two roughly equal ranges in metallicity using the metallicity gradient measurements from \citet{mou10}.

We determine the average HI depletion times by summing up the HI column densities and the SFRs from all regions under consideration.
The errorbars shown in Figure~\ref{fig:met} incorporate all uncertainties related to HI, UV and IR flux measurements as well as the uncertainty in SFR calibration.
For the THINGS galaxies we see that when radial distance from the galaxy centre is used as a proxy for metallicity of the ISM, the HI depletion time increases with decreasing metallicity within spiral galaxies.  
But we also derive the average depletion times for all HI-dominated regions of THINGS galaxies, which is a more appropriate measure to compare with the HI depletion time derived for FIGGS galaxies (which do not have metallicity gradient measurements).
And on comparing the average values we notice that the HI depletion times are remarkably uniform across the large range in metallicity sampled by dwarf irregulars and spirals.
We have checked that THINGS regions mapped at 1 kpc resolution, and `Bluedisk' control galaxies give comparable values for the measured average HI depletion time.

The amount of dust in the ISM scales with its measured metallicity.
(i) The rate of H$_{\rm 2}$ formation and (ii) the shielding of UV radiation enabling the physical conditions for H$_{\rm 2}$ formation, are both proportional to the amount of dust in the ISM, and are therefore expected to be proportional to the metallicity of the ambient ISM \citep{kru09a,ste14}.
There also exist observational evidence of the dependence of H$_{\rm 2}$-to-HI conversion on metallicity \citep[e.g.][]{wel12,won13}.
The H$_{\rm 2}$ thus formed is expected to be the phase correlated with star formation as in this phase cooling and fragmentation of gas proceeds very efficiently \citep[e.g.][]{kru11}, and as mentioned previously observations seem to confirm this in the central regions of spirals \citep{ler13}.
One would thus expect that a higher efficiency of conversion of HI to H$_{\rm 2}$ should reflect in a lower gas depletion time / more efficient Kennicutt-Schmidt relation. 
That is why our finding that the gas depletion time / Kennicutt--Schmidt relation is independent of the metallicity in HI dominated regions of star forming galaxies is surprising.
Though it must be noted that molecular gas formation might not be determined by metallicity alone, and factors like the local pressure and volume densities could also play a role.

\subsection{Comparison with theoretical models}

\begin{figure*}
\begin{center}
\begin{tabular}{ccc}
{\mbox{\includegraphics[width=5.7cm]{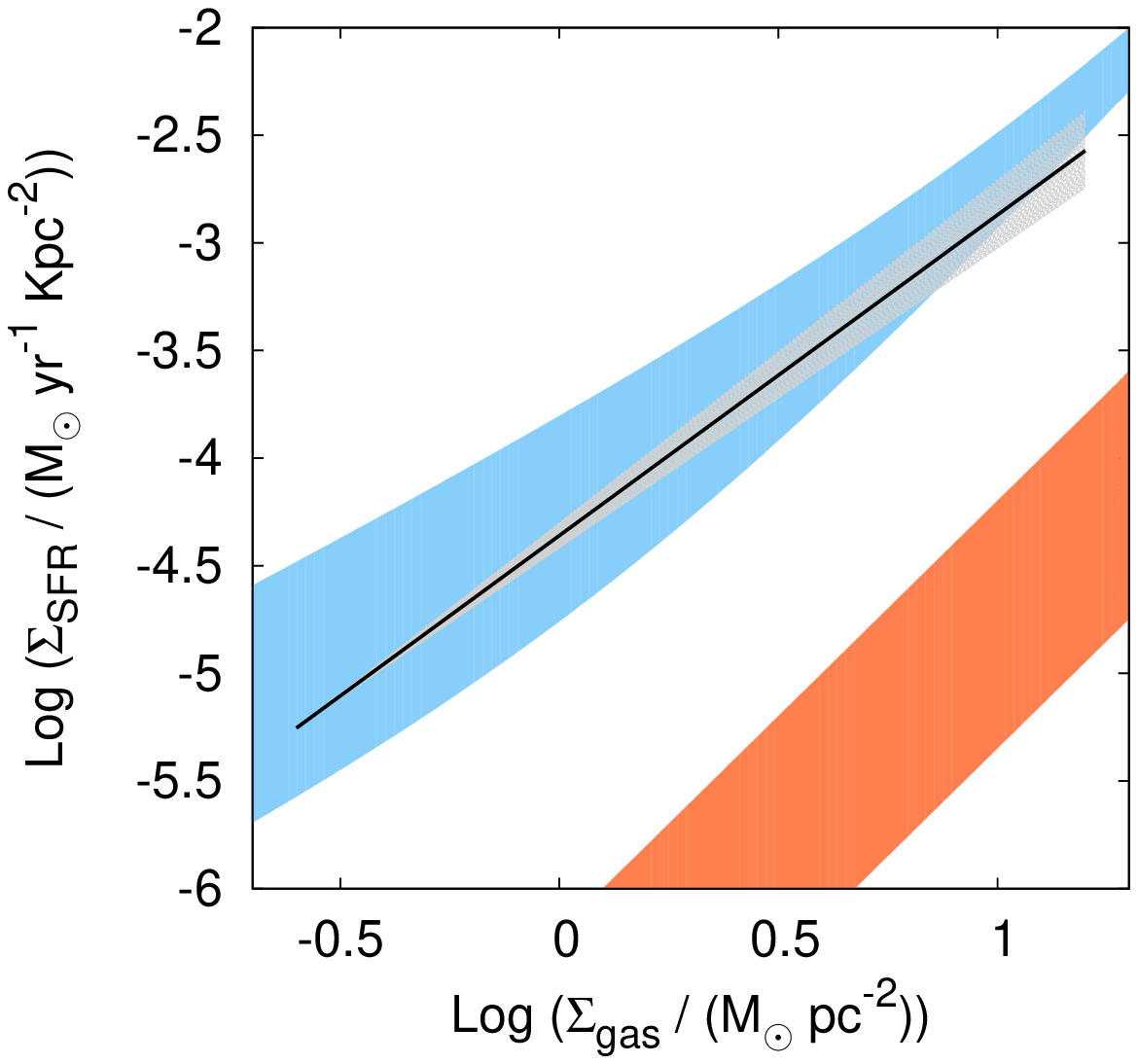}}}&
{\mbox{\includegraphics[width=5.7cm]{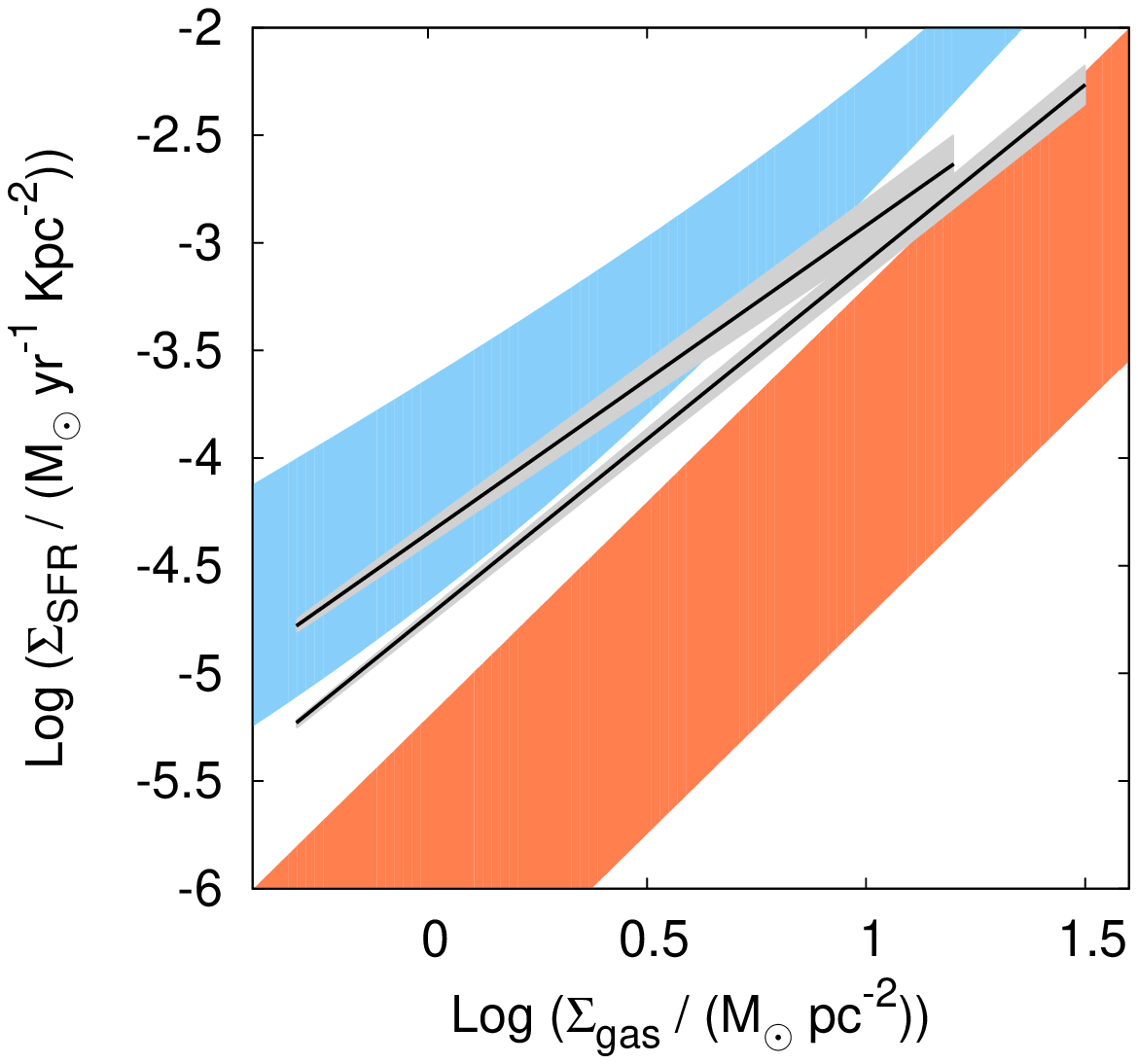}}}&
{\mbox{\includegraphics[width=5.7cm]{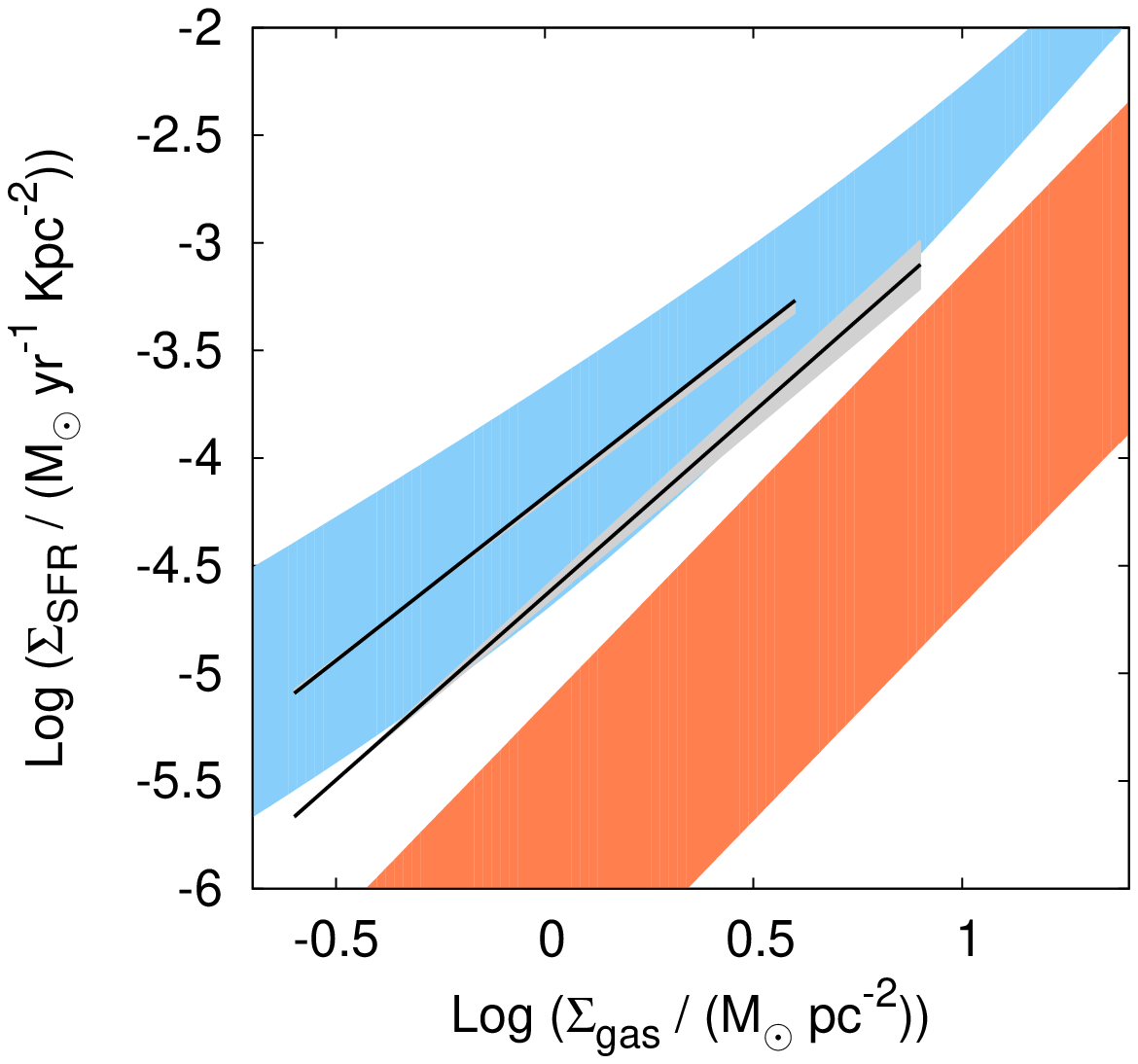}}}\\
\end{tabular}
\caption{Comparison of the K--S relations determined by us for FIGGS dwarfs (left panel), THINGS spirals (middle panel) and `Bluedisk' galaxies (right panel) with OML (blue shaded area) and KMT (red shaded area) models. See text for details of model parameters used. 1 $\sigma$ errors on the fits are shown in grey.}
\label{fig:th}
\end{center}
\end{figure*}

In this section, we compare our derived  K--S relations with predicted trends 
from two widely used models of star formation based on the following papers: 
(i) \citet{ost10} hereafter OML, (ii) \citet{kru09} hereafter KMT.
The OML model assumes that atomic gas has achieved two-phase thermal and quasi-hydrostatic equilibrium.
In the KMT model, hydrostatic balance sets the requirements for the 
formation of the CNM phase and the atomic-to-molecular transition, 
except at very low metallicities where it is set by the two-phase 
thermal equilibrium at the maximum allowed CNM temperature \citep{kru13}.
We consider the predictions of the respective models for the low SFR, low metallicity, HI-dominated regime.
Therefore for OML model we consider equation (22) from \citet{ost10}.
For the KMT model, we consider the combination of equations (18) and (20) from \citet{kru13}.

When comparing the models with our observations four parameters used in both the OML and KMT models need to be considered: metallicity, $\rho_{sd}$, $f_w$ and the clumping factor.
For the FIGGS galaxies the metallicity is fixed at 0.1 Z$_{\odot}$, 
whereas for the spirals from THINGS and `Bluedisks' surveys, the metallicity is varied between 0.4 to 1 Z$_{\odot}$.
The parameter $\rho_{sd}$ in the models which refers to the density of stellar $+$ dark matter is varied between 0.003 to 0.03 ${\rm M_{\odot}~pc^{-3}}$ \citep[following][]{kim11}.
The parameter $f_w$ is the models which is related to the fraction of neutral 
gas in the warm phase is varied between 0.05 and  1.
From \citet{ler13b} we know that HI clumps much less when observed at high resolution compared to CO, and in fact the clumping factor for HI remains almost constant.
As  we are only sensitive to HI emission, we follow the empirical relation given in \citet{ler13b} and
adopt a clumping factor of $\sim$ 1.3 at 400 pc and 
1 kpc scales (in comparison with what will be observed at 100 pc resolution) and $\sim$ 1.5 at 10 kpc scales.

Models with the range in parameters given above are shown in Figure~\ref{fig:th}.
We see that the predictions of the OML model match our observations better than the predictions of the KMT model for galaxies of all types.
It is interesting that our data agrees with the OML model, since recent hydrodynamical simulations \citep{kim11,kim13} have shown that turbulence 
driven by supernova feedback regulates the thermal pressure in the outer discs of galaxies, 
and this thermal pressure plays a crucial role in the OML models.
This may imply that the ISM in outer discs of spirals and dwarfs is stabilized by feedback.
Recent hydrodynamical simulations of galaxy formation have emphasized the 
 primacy of feedback in determining how gas converts into stars \citep[e.g. see][]{hop14}.

\section{Summary}

We study the spatially resolved Kennicutt--Schmidt relation between \ssfr\ and \shi\ 
in the HI dominated ISM in spiral and dwarf irregular galaxies across a wide range of spatial scales
and gas surface densities.
We account for the uncertainties associated with measuring low SFRs within small regions by averaging over
SFR in fixed bins of HI surface density.
For dwarfs to spirals, and across different scales, the K--S relation is found to have 
a slope of $\sim$1.5. The efficiency with which the cold neutral gas is being
converted into starts is  an order of magnitude more inefficient than within the optical discs of spirals. 
A surprising conclusion from our findings in that the mean Kennicutt-Schmidt relation in HI dominated regions is independent of metallicity.
Our results favour a model of star formation where thermal and 
pressure equilibrium in the outer ISM regulate the rate at which star formation occurs, where the thermal pressure in turn is set by supernova feedback.

\section*{Acknowledgments}
We would like to thank the referee, Frank Bigiel for his constructive comments which helped improve the paper.
We thank Ayesha Begum for providing the visibilities which were used to derive the HI maps for FIGGS galaxies used in this work.
Some of the data presented in this paper were obtained from the Mikulski Archive for Space Telescopes (MAST). STScI is operated by the Association of Universities for Research in Astronomy, Inc., under NASA contract NAS5-26555. Support for MAST for non-HST data is provided by the NASA Office of Space Science via grant NNX13AC07G and by other grants and contracts.
This research has made use of the NASA/ IPAC Infrared Science Archive, which is operated by the Jet Propulsion Laboratory, California Institute of Technology, under contract with the National Aeronautics and Space Administration.
This publication makes use of data products from the Wide-field Infrared Survey Explorer, which is a joint project of the University of California, Los Angeles, and the Jet Propulsion Laboratory/California Institute of Technology, funded by the National Aeronautics and Space Administration.

\bsp

\label{lastpage}

\end{document}